# A Novel Approach for Imputation of Missing Attribute Values for Efficient Mining of Medical Datasets – Class Based Cluster Approach

Y.Usha Rani[1], P. Sammulal[2]

[1] Faculty of Information Technology, VNR Vignana Jyothi Institute of Engineering and Technology, Hyderabad, 500090, India
[2] Professor, Department of Computer Science and Engineering, JNT University, Hyderabad, 500090, India
[1] Email :usharani_y@vnrvjiet.in

**Abstract.** Missing attribute values are quite common in the datasets available in the literature. Missing values are also possible because all attributes values may not be recorded and hence unavailable due to several practical reasons. For all these one must fix missing attribute vales if the analysis has to be done. Imputation is the first step in analyzing medical datasets. Hence this has achieved significant contribution from several medical domain researchers. Several data mining researchers have proposed various methods and approaches to impute missing values. However very few of them concentrate on dimensionality reduction. In this paper, we discuss a novel imputation framework for missing values imputation. Our approach of filling missing values is rooted on class based clustering approach and essentially aims at medical records dimensionality reduction. We use these dimensionality records for carrying prediction and classification analysis. A case study is discussed which shows how imputation is performed using proposed method.

## 1. INTRODUCTION

Preprocessing medical records is one of the important steps one cannot avoid when handling medical datasets. This is because different attributes may be having different attribute value types and must be taken into consideration to understand and handle them appropriately. Also values of attributes are of most important concern when imputing missing values and performing classification. For example, if we choose to adopt Euclidean distance measure without performing normalization of attribute values then an attribute having maximum value may affect the distance computations which may cause wrong classification and imputations. In short, attribute domain and range requires to be considered to handle them and also use of appropriate distance measure. Also, the diverse nature of medical records makes handling medical records quite challenging for data analysts and researchers. Some important challenges when handling medical records includes applying smoothing methods to medical records, discovering outliers in medical data, estimation of class labels and imputing missing values, normalizing medical attributes, handling inconsistent medical data.

Data preprocessing techniques affect Data Quality. Efficient preprocessing improves data quality. In this sense, data preprocessing approaches gained major importance from data analysts and miners. Also, improper data values and incorrect data values mislead both the prediction and classification results, which further results in false classification results. This in turn leads to improper medical treatment which is a very dangerous potential hazard. The present research focus is mainly to effectively handle missing attribute values present in medical records of a dataset. The attributes may be categorical, numeric. Our method can handle all attribute types without the need to devise a different method to handle different attribute types. This is first importance of our approach. We must normalize data as different attributes may have different ranges. We outline research objective and problem specification in the succeeding lines of this paper and then discuss importance of our approach.

### a. Research Aim

The contribution for the present research includes following objectives
  i.   The first objective is to fix missing values.
  ii.  The second objective is to reduce dimensionality of medical records, without losing important information.
  iii. Perform classification of medical records
  iv.  Predict the best possible imputation value

### b. Problem Definition

To use the concept of clustering for dimensionality reduction of medical records and perform imputation using proposed class based clustering approach of dimensionality reduction and finally classify the record label of a medical record and also predict the disease severity if known disease levels are available as part of dataset.

*c.  Significance of Proposed Method*
i)   Using the proposed approach we may impute both categorical and numerical attribute values
ii)  The approach is extended for finding class labels of medical records
iii) Disease prediction using the proposed approach

One interesting paper in literature which motivates and debates whether missing values are important or not important. This question is addressed in research of (Zhang, S, 2005). The imputation based on clustering is addressed in (Zhang, C, 2006) and (Wang, Ling, 2010) uses concept of support vector regression to imputation. (Zhenxing Qin , 2006) discuss the problems if missing values are to be imputed or discarded and also throw light on decision tree cost incurred.( Farhangfar A,2007) address a novel imputation framework and methodology.  (Jau-Huei Lin, 2008) research is concerned with using Bayesian network for estimation of clinical missing values.( Miew Keen Choong,2009) work is based on auto regression concept.(Kirkpatrick B,2010) discuss Phylogeny problems in missing values. (Shobeir Fakhraei, 2010) use consensus approach for classifying bio-medical datasets. (Xiaofeng Zhu, 2011) address missing values problem in datasets having mixed attributes. They use concept of feature ranking. A new and efficient approach for detecting intrusion is discussed in research work of authors (Wei-Chao Lin, 2015). Our imputation framework is motivated from this research. However, the present imputation approach proposes class-based imputation strategy to estimate missing value(s) in records.

## 2. RESEARCH ISSUES IN MINING MEDICAL DATA

### 2.1 Research Issue 1: Datasets

Handling Datasets requires knowledge about preprocessing techniques. A dataset cannot be handled directly and cannot be taken granted to be used directly. Sometimes we may have to perform normalization of medical records of datasets considered so that they become feasible to be used by the proposed algorithm or methodology. Medical datasets have finite and limited set of attributes but requires proper handling which otherwise leads to failure in classification or incorrect disease prediction.

### 2.2 Research Issue 2: Missing Values

Missing values are quite common in medical records. This happens because of several reasons. Many times this is due to unrecorded tests as required tests are not conducted. Sometimes the values of all test attributes may not be collected due to several legitimate reasons. Whatever may be the reason for missing values, we cannot take the comfort of mis-handling of medical records as accuracy cannot be the relaxing factor in disease prediction. In literature several researchers deal the method of missing values as discussed in section.1, however in this paper we aim class based clustering imputation strategy to fix missing values. Our approach fixes both categorical and numerical attributes values and we discuss generalized approach for handling missing values in the sections below.

### 2.3 Research Issue 3: Prediction or Classifications algorithms

Predicting disease through imputing values of medical records is challenging compared to performing supervised classification. This is because any inefficiency in handling missing values may land classification or prediction algorithms in to wrong prediction. To date, though there are several classification accuracy estimation procedures, there is no single solution which may be treated to be best approach.

### 2.4 Research Issue 4: Nearest Record and Class Label

The method adopted to impute missing attribute values decides performance of classification algorithm or classifiers. In this context, the approach followed for imputation gains its importance. Also, the classification accuracy depends on the accuracy of distance given by distance measure used. The class label of record is usually the one for which test record distance is minimum. Usually, we may use KNN, SVM classifier for defining class labels.

### 2.5 Research Issue 5 : Medical Parameters

Classification accuracy is a function of dominant medical attributes. Information of these medical attributes decides the classification accuracy. Approaches for deciding the most dominant attributes are the deciding factors for evaluating classification accuracy.

**2.6 Research Issue 6: Noise or Outlier Attributes**

Attributes with low significance and less dominant may be neglected in evaluating classification accuracies. The attributes we eliminate are called outlier attributes. Sometimes, such outliers may also affect classification accuracy in negative sense. A proper care must be taken while eliminating outlier attributes.

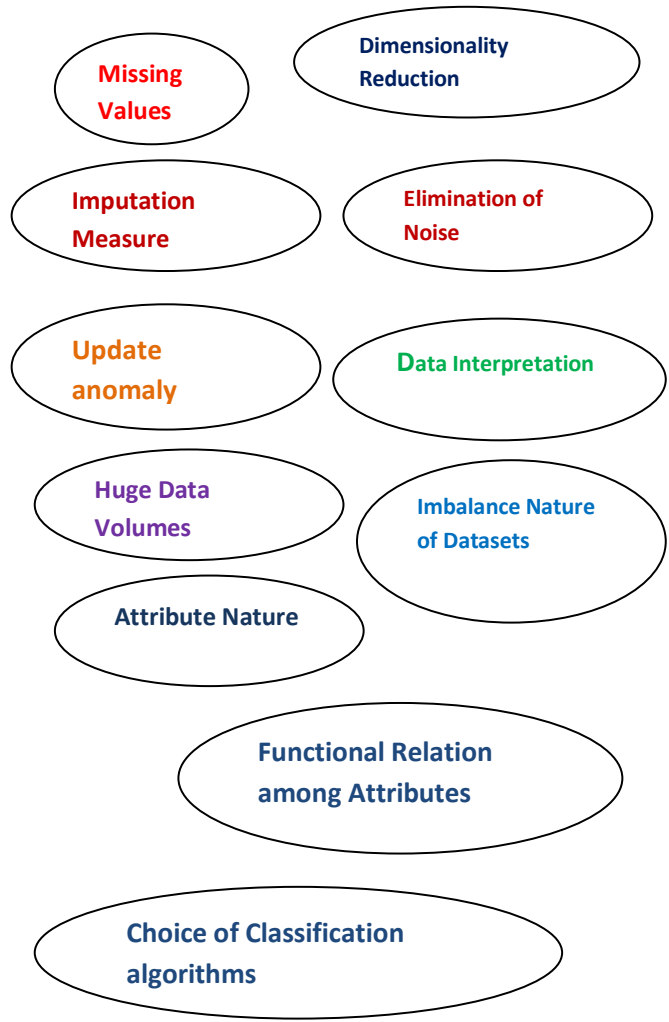

**Fig.1 Research Problems when handling Medical Datasets**

## 3. IMPUTATION FRAMEWORK

In this section, the framework for our proposed imputation approach is discussed. Initially we group records into two different groups, those without missing attribute values ($G_1$) and another group ($G_2$) having missing attribute values. The idea is to consider all records in group $G_1$ (having no missing values) and first obtain clusters equal to number of decision labels and use knowledge of cluster information to achieve dimensionality reduction and imputation. From clusters obtained considering records in first group ($G_1$) and second group ($G_2$) using any one of the existing clustering algorithms, we obtain cluster mean for all clusters. We also obtain nearest neighbors for each record equal to number of decision classes. Each record now is transformed to a distance value obtained by summing distance of each record to its cluster centers and distance of first two nearest neighbors for each record.

We call this distance as mapping distance. This mapping distance is the base for deciding imputation. The following steps give outline of proposed framework.

*a. Obtain Clusters from medical records of first Group $G_1$*

The process of imputation starts with considering records with no missing values and forming clusters from these records. Then, we obtain mean for all clusters.

*b. Computing distance (type-1) of normal records to Cluster Centers*

Considering each cluster mean, we obtain distance of each medical record to cluster mean. Then, we obtain sum of all these distances computed. Now, all records are represented as distance values of single dimension. Call this distance as distance type-1.

*c. Computing distance (type-2) of missing records to Cluster Centers*

Consider each record in $G_2$; obtain distance between all these records w.r.t each cluster mean. When finding distance value, we must consider only those attributes which do not have missing values. We then add all these distances. Call this distance as distance type-2.

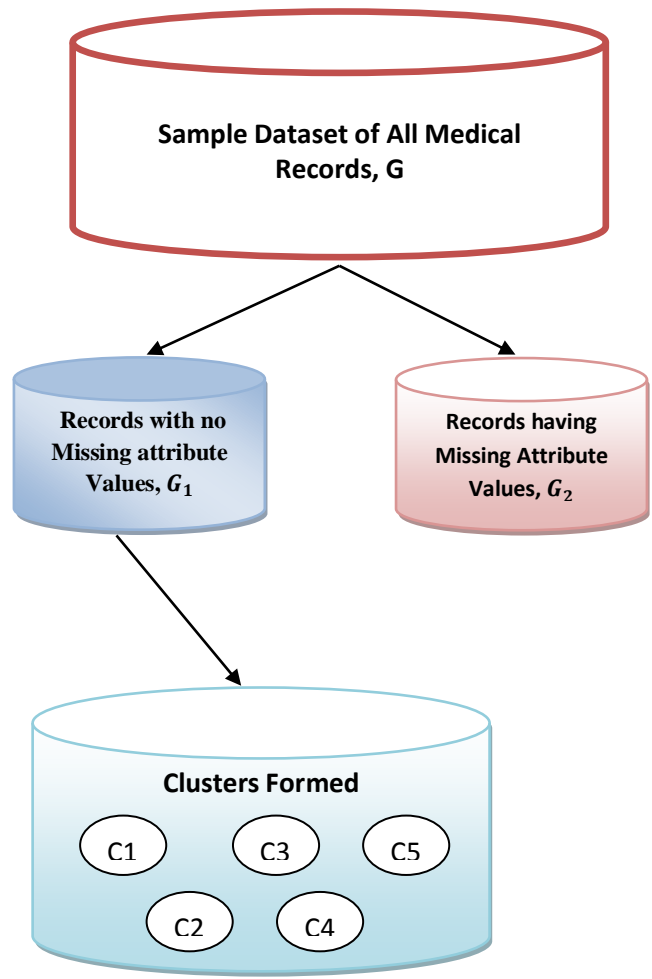

Fig 2. Outline of Procedure for Generating Clusters

*d. Obtain Nearest Neighbors*

Consider each record in $G_1$ and obtain first, 'd', intra-cluster nearest neighbors. Here, d, is number of decision labels. Sum all these distances to record distances obtained in step-b respectively. We then find the distance between each record in $G_2$ and $G_1$. From these distance values, choose only first 'd' minimum distances. Sum all these distances to record distances (type-2 distance) obtained in step-c respectively. In this paper, we term these distances obtained finally as mapping distances.

## 3.1 PROPOSED IMPUTATION METHOD

*Input*: *Records with and without Missing Values*

*Output*: *Missing Values Imputed*

*Notations adopted* :

$R_i$ — $i^{th}$ *medical record*
$R_i(A_K)$ — $k^{th}$ *attribute value of* $i^{th}$ *medical record*
$G_c$ — $c^{th}$ *group*
$i, k$ — *index of medical records and attributes*
$\emptyset$ — *misisng record or Empty record value*
$c$ — *number of decision classes in medical dataset*
$D_d$ — $d^{th}$ *decision class*
$m$ — *total number of medical records*
$n$ — *number of attributes in each record*
$\mu_d$ — *cluster center of* $d^{th}$ *cluster*
$\mu_{dn}$ — *mean value of* $n^{th}$ *attribute*
$h$ — *number of records in group* , $G_2$
$z$ — *number of records in group* , $G_1$ *equal to* $(m - h)$

**Step-1: Scan Dataset of Medical Records**

We start with scanning all medical records and classify them into two categories. The first group includes all medical records which do not have missing values and second group contains medical records with missing values. The former is denoted as group, $G_1$ and later is denoted as group $G_2$. We denote these two groups formally using equations defined in equations (1) and (2) respectively.

$$G_1 = U \{ R_i \mid R_i(A_K) \neq \emptyset , \forall\, i, k \} \qquad (1)$$

and

$$G_2 = U \{ R_i \mid R_i(A_K) = \emptyset \ / \exists\, i, k \} \qquad (2)$$

where $\emptyset$, denotes undefined (empty or missing values), i denote indices of medical record with $i \in (1, m - h)$ and k denotes attribute of feature set with $k \in (1, n)$. In our case, group, $G_1$, of medical records is considered as training set and all medical records in group, $G_2$ is considered to be testing test. The medical records present in the group, $G_1$, are used to build the knowledge base. It is this knowledge base, over which we test the medical records in group, $G_2$.

**Step-2: Cluster Medical records in group, $G_1$**

After classifying medical records in to two groups, $G_1$ and $G_2$ , the objective is to obtain maximum number of decision classes. Let, g = |$D_d$ |, be total number of decision classes. Cluster all those medical records in group, $G_1$ to a number of clusters equal to |$D_d$|. To obtain predefined number of clusters, we may use k-means clustering algorithm. The value of K is |$D_d$| in our case. The main reason for obtaining clusters is dimensionality reduction. Here we choose to transform records of higher dimensions to their equivalent lower dimension. At the end of this step, we obtain clusters consisting of medical records from group, $G_1$ as shown in figure-2.

**Step-3: Determine mean vector of each cluster**

After step-2, we have clusters formed. After obtaining the clusters formed using k-means approach, we find mean vector of each cluster. The mean vector of each cluster is obtained by obtaining mean of each attribute. Let us suppose, the cluster represented as $C_d$ denotes $d^{th}$ - cluster consisting four records namely $R_1$, $R_5$, $R_7$, and $R_9$ with only a single attribute say, $A_1$. If every record has single attribute, then cluster mean is given by equation (3)

$$\mu_d = \frac{R_1(A_1) + R_5(A_1) + R_7(A_1) + R_9(A_1)}{4} \tag{3}$$

In short, cluster mean of $g^{th}$ cluster is computed using generalized equation (4)

$$\mu_g = U_k \left[ \frac{\{\sum R_l^k \mid l \, \epsilon \, \{1, q\} \, for \, each \, k \, \epsilon \, \{1, n\} \}}{|l|} \right] \tag{4}$$

In this paper, we use notation, $U_k$ to represent union of all values each separated by a symbol comma. Each cluster center, $\mu_g$, is given by equation (5)

$$\mu_g = (\mu_g^1, \mu_g^2, \mu_g^3, \ldots \ldots \mu_g^n) \tag{5}$$

Here '$\mu_g$ is a sequence of 'n' values denoting $g^{th}$ cluster mean over 'n' attributes and $\mu_g^i$ indicates mean of $i^{th}$ attribute in $g^{th}$ cluster. The value of 'n' represents total number of attributes; |g| indicates number of clusters.

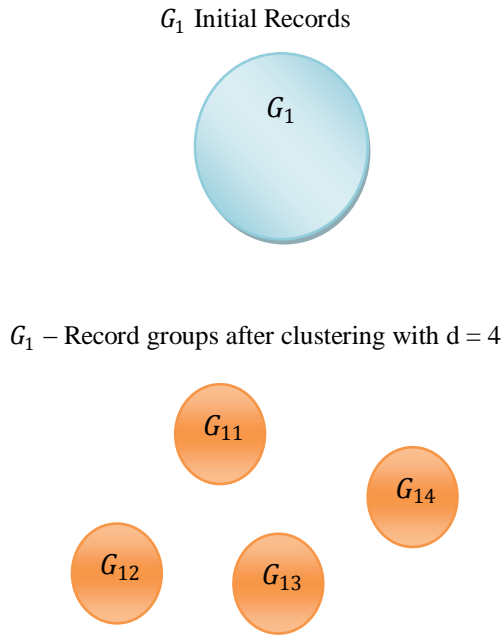

$G_1$ Initial Records

$G_1$ – Record groups after clustering with d = 4

Fig 3. Clustering Records Based on Number of Decision Labels

**Step-4: Compute Type-1 distance between each record, $R_i$ and cluster centers**

For clusters obtained, compute distance of every record to mean of every cluster. This is done by finding Euclidean distance between each medical record with 'n' attributes and mean vector of every cluster. These distances are all summed to obtain initial representative of each record. This distance representative for each record is termed as Type-1 distance of that record respectively and is given by equation 6 below.

$$Dist^1(\boldsymbol{R_i}, \boldsymbol{\mu_d}) = \sum_{d=1}^{d=|g|} Dist^d(R_i, \mu_d) \tag{6}$$

With

$$Dist^d(R_i, \mu_d) = \sqrt{(R_{i1} - \mu_d^1)^2 + (R_{i2} - \mu_d^2)^2 + \cdots (R_{in} - \mu_d^n)^2} \quad \boldsymbol{for} \; i \, \epsilon \, (1,n), \forall d \tag{7}$$

At the outset of this step, we have distance value from record , $R_i$ to each cluster mean represented by $\mu_d$.

**Step-5: Obtain |m|-nearest neighbors and use mapping function to transform multi-dimensional record to a single value (one dimension representative)**

Let $Map(R_i)$ be the mapping function which maps a record, $R_i$ on to a 1D equivalent distance value and $NN_d$ be the nearest neighbor. Alternately, it may also indicate similarity computation. In such a case, we consider the maximum value. To determine mapping function value of a record we use the equation (8)

$$Map(R_i) = Dist^1(\boldsymbol{R_i}, \boldsymbol{\mu_d}) + distance \; of \; r - nearest \; neighbors \quad \forall \, i \, \epsilon \, (1, n-h) \tag{8}$$

With |g| denotes number of clusters formed, r varies from 1 to |d| and |d| is number of decision classes and (n-h) indicates number of records in group, $G_1$.

The output of step-5, is that each medical record, $R_i$ is finally mapped to a single dimension. In other words, the medical record of 'n' dimensions is reduced to single dimension achieving dimensionality reduction.

**Step-6: Compute Type-2 distances between each cluster mean and missing attribute value record, $\boldsymbol{R_j}$ in group, $\boldsymbol{G_2}$**

As in step-4, we find distance between each record with missing attribute value and each cluster mean. When finding Type-2 distance, we discard missing attribute value ($y^{th}$) and corresponding $y^{th}$ value in mean vector. All these distances are summed to obtain single distance value. This distance is called Type-2 distance value and may be obtained using equations (9) and (10).

$$Dist^2(\boldsymbol{R_j}, \boldsymbol{\mu_d}) = \sum_{d=1}^{d=|g|} Dist^d(\boldsymbol{R_j}, \boldsymbol{\mu_d}) \tag{9}$$

Where

$$Dist^d(\boldsymbol{R_{jn}}, \boldsymbol{\mu_{dn}}) = \sqrt{\sum(R_{j1} - \mu_{d1})^2 \ldots \ldots \ldots} \; \forall R_{jn} \; where \; n \neq y \tag{10}$$

**Step-7: Obtain |m|-nearest neighbors and use mapping function to transform missing attribute value record to a single value**

$Map^r(R_j)$ is function which maps record, $R_j$ to a single distance value. To determine mapping function value of a record $R_j$, we use the equation.11

$$Map^r(R_j) = Dist^2(\boldsymbol{R_j}, \boldsymbol{\mu_d}) + distance \; of \; r - nearest \; neighbors \quad \forall \, j \, \epsilon \, (1, h) \tag{11}$$

With |g| is number of clusters formed and $j\epsilon(1,h)$

The output of step-7 is that each record $R_j$ is mapped to low dimension.

**Step-8: Compute distances between mapping distances obtained in step-5 and step-7**

For all records denoted by $R_j$ in $G_2$, obtain difference between mapping distance of every record, $R_i$ in group, $G_1$ and record for which missing value must be imputed say, $R_j$ in group, $G_2$. This difference is denoted by $d_{ij}$.

**Step-9: Find nearest record for achieving imputation**

$R_j$ is considered as similar $R_i$ whose corresponding $d_{ij}$ is most minimum. This is given by equation. (12)

$$d_{ij} = | Map(R_i) - Map^r(R_j) | \mid_{\min} \tag{12}$$

**Step-10: Estimate Missing values and Impute**

Impute missing attribute value of record, $R_j$ denoted by $R_{jr}$ by the attribute value, $R_{ir}$ of medical record denoted as $R_i$. When we consider top-k minimum values, impute missing attribute value by considering decision class whose frequency is maximum. Also, when imputing, we may impute the missing attribute value with the mean attribute value of the records considered as top-k.

## 4. CASE STUDY

In this Section, we discuss the proposed methodology for imputing missing attributes values by considering sample data of medical records as given in Table. I. The sample consists both categorical and numerical attribute values. Table. II represents normalized records. Table.III represents medical records which are free from missing values and also those records which have missing values.

**TABLE I. INPUT DATA OF MEDICAL RECORDS**

| Medical Records | Attributes | | | | Class Label |
|---|---|---|---|---|---|
| | Z1 | Z2 | Z3 | Z4 | |
| MR1 | $K_{11}$ | 5 | $J_{31}$ | 10 | C-1 |
| MR2 | $K_{13}$ | 7 | $J_{31}$ | 5 | C-1 |
| MR3 | $K_{11}$ | 7 | $J_{32}$ | 7 | C-1 |
| MR4 | $K_{12}$ | 5 | $J_{31}$ | 10 | C-1 |
| MR5 | $K_{13}$ | 3 | $J_{32}$ | 7 | C-2 |
| MR6 | $K_{12}$ | 9 | $J_{31}$ | 10 | C-2 |
| MR7 | $K_{11}$ | 5 | $J_{32}$ | 3 | C-2 |
| MR8 | $K_{13}$ | 6 | $J_{32}$ | 7 | C-2 |
| MR9 | $K_{12}$ | 6 | $J_{32}$ | 10 | C-2 |

**TABLE II. NORMALIZED REPRESENTATION**

| Medical Records | Attributes | | | | Class Label |
|---|---|---|---|---|---|
| | Z1 | Z2 | Z3 | Z4 | |
| MR1 | 1 | 5 | 1 | 10 | C-1 |
| MR2 | 3 | 7 | 1 | 5 | C-1 |
| MR3 | 1 | 7 | 2 | 7 | C-1 |
| MR4 | 2 | 5 | 1 | 10 | C-1 |
| MR5 | 3 | 3 | 2 | 7 | C-2 |
| MR6 | 2 | 9 | 1 | 10 | C-2 |
| MR7 | 1 | 5 | 2 | 3 | C-2 |
| MR8 | 3 | 6 | 2 | 7 | C-2 |
| MR9 | 2 | 6 | 2 | 10 | C-2 |

**TABLE III. DATASET WITH AND WITHOUT MISSING ATTRIBUTE VALUES**

| Medical Records | Attributes | | | | Class Label |
|---|---|---|---|---|---|
| | Z1 | Z2 | Z3 | Z4 | |
| MR1 | 1 | 5 | 1 | 10 | C-1 |
| MR2 | 3 | 7 | 1 | 5 | C-1 |
| MR3 | 1 | 7 | ? | 7 | C-1 |
| MR4 | 2 | 5 | 1 | 10 | C-1 |
| MR5 | 3 | 3 | 2 | ? | C-2 |
| MR6 | 2 | 9 | 1 | 10 | C-2 |
| MR7 | 1 | 5 | 2 | 3 | C-2 |
| MR8 | 3 | 6 | 2 | 7 | C-2 |
| MR9 | 2 | 6 | 2 | 10 | C-2 |

Table IV denotes all records without missing values while Table. V shows records with missing attribute values. Table.VI below shows clusters obtained considering medical records in group $G_1$, by considering k=2 in k-means algorithm. The clusters obtained using k-means algorithm is namely $C_1$ and $C_2$.

**TABLE IV. RECORDS WITHOUT MISSING VALUES**

| Medical Record | \_ Attributes \_ | | | | Class Label |
|---|---|---|---|---|---|
| | Z1 | Z2 | Z3 | Z4 | |
| MR1 | 1 | 5 | 1 | 10 | CLASS-1 |
| MR2 | 3 | 7 | 1 | 5 | CLASS-1 |
| MR4 | 2 | 5 | 1 | 10 | CLASS-1 |
| MR6 | 2 | 9 | 1 | 10 | CLASS-2 |
| MR7 | 1 | 5 | 2 | 3 | CLASS-2 |
| MR8 | 3 | 6 | 2 | 7 | CLASS-2 |
| MR9 | 2 | 6 | 2 | 10 | CLASS-2 |

**TABLE V. RECORDS WITH MISSING VALUES**

| Medical Record | \_ Attributes \_ | | | | Class Label |
|---|---|---|---|---|---|
| | Z1 | Z2 | Z3 | Z4 | |
| MR3 | 1 | 7 | 2 | 7 | CLASS-1 |
| MR5 | 3 | 3 | 2 | 7 | CLASS-2 |

**TABLE VI. CLUSTERS GENERATED FROM TABLE.III**

| Generated CLUSTERS | RECORDS |
|---|---|
| CLUSTER.1 | MR1,MR4,MR6,MR9 |
| CLUSTER.2 | MR2,MR7,MR8 |

**TABLE VII. CLUSTERS WITH MEANS**

| CLUSTERS | Mean Attribute Values | | | |
|---|---|---|---|---|
| | Z1 | Z2 | Z3 | Z4 |
| CLUSTER-1 | 2.33 | 6 | 1.67 | 5 |
| CLUSTER-2 | 1.75 | 6.25 | 1.25 | 10 |

The first cluster $C_1$ consists set of all medical records {$MR_1$, MR4, $MR_6$, $MR_9$} and second cluster $C_2$ consists of set of all medical records {$MR_2$, $MR_7$, $MR_8$ }. The cluster mean is represented in Table.VII. The distance of records to mean vector of cluster-1 and cluster-2 is denoted in Table. VIII and Table. IX respectively. Type-1 distance of records computed is denoted in Table.X.

**TABLE VIII. DISTANCE OF RECORDS TO CLUSTER-1**

| Medical Records | Distance to first cluster |
|---|---|
| MR1 | 5.312459 |
| MR2 | 1.374369 |
| MR4 | 5.153208 |
| MR6 | 5.878397 |
| MR7 | 2.624669 |
| MR8 | 2.134375 |
| MR9 | 5.022173 |

**TABLE IX. DISTANCE OF RECORDS TO CLUSTER-2**

| Medical Records | Distance to second cluster |
|---|---|
| MR1 | 1.47902 |
| MR2 | 5.214163 |
| MR4 | 1.299038 |
| MR6 | 2.772634 |
| MR7 | 7.189402 |
| MR8 | 3.344772 |
| MR9 | 0.829156 |

**TABLE X. Type-1 DISTANCE OF GROUP-1 RECORDS**

| Medical Records | Type-1 distance computation |
|---|---|
| MR1 | 6.791479 |
| MR2 | 6.588532 |
| MR4 | 6.452246 |
| MR6 | 8.651031 |
| MR7 | 9.814071 |
| MR8 | 5.479147 |
| MR9 | 5.851329 |

The record distances in group, $G_2$ to clusters means and type-2 distance computations are shown in Table.XI and Table.XII respectively.

**TABLE XI. DISTANCE OF RECORDS IN G2 TO CLUSTERS FORMED**

| Medical Records | Distance to first cluster | Distance to second cluster |
|---|---|---|
| MR3 | 2.603417 | 3.181981 |
| MR5 | 3.091206 | 3.561952 |

**TABLE XII.  Type-2 DISTANCE OF GROUP-2 RECORDS**

| Record | Type-2 distance computation |
|---|---|
| MR3 | 6.791479 |
| MR5 | 6.588532 |

The pair wise distance computations of medical records in group G1 and G2 are shown in Table.XIII and Table.XIV respectively. The mapping distance of medical records of group1 considering the nearest neighbor computations is given in Table.XV.The mapping distance is sum of initial mapping distances type-1 and type-2 (for missing value records) and nearest neighbor distance.

**TABLE XIII. GROUP-1 PAIRWISE DISTANCE COMPUTATION**

|  | MR1 | MR4 | MR6 | MR9 |
|---|---|---|---|---|
| MR1 | 0 | 1 | 4.123106 | 1.732051 |
| MR4 | 1 | 0 | 4 | 1.414214 |
| MR6 | 4.1231 | 4 | 0 | 3.162278 |
| MR9 | 1.732 | 1.414 | 3.16 | 0 |

**TABLE XIV. GROUP-2 PAIRWISE DISTANCE COMPUTATION**

|  | MR2 | MR7 | MR8 |
|---|---|---|---|
| MR2 | 0 | 3.605551 | 2.44949 |
| MR7 | 3.605551 | 0 | 4.582576 |
| MR8 | 2.44949 | 4.582576 | 0 |

**TABLE XV.  FINAL MAPPING DISTANCE OF GROUP-1 RECORDS**

| Record | Distance To Clusters | First Nearest Neighbor Distance | Second Nearest Neighbor Distance | Final Mapping Distance |
|---|---|---|---|---|
| MR1 | 6.791479 | 1 | 1.732051 | 9.52353 |
| MR2 | 6.588532 | 2.44949 | 3.605551 | 12.64357 |
| MR4 | 6.452246 | 1 | 1.414214 | 8.86646 |
| MR6 | 8.651031 | 3.16 | 4 | 15.81331 |
| MR7 | 9.814071 | 3.605551 | 4.582576 | 18.0022 |
| MR8 | 5.479147 | 2.44949 | 4.582576 | 12.51121 |
| MR9 | 5.851329 | 1.414 | 1.732 | 8.997329 |

Table.XVI shows computations of distances between group $G_1$ medical records and medical record $MR_3$. Similarly Table.XVII shows computations of distances between group $G_1$ medical records and medical record $MR_5$. From Table.XVI and Table.XVII, we can depict that the records $R_3$ and $R_5$ are similar to record $R_8$ and hence the missing values may be imputed accordingly.

**TABLE XVI   DISTANCE BETWEEN MAPPING DISTANCE OF GROUP-1 MEDICAL RECORDS AND MR3**

| Records | Mapping Distance | Record | Mapping Distance | Distance |
|---------|-----------------|--------|-----------------|----------|
| MR1 | 9.52353 | MR3 | 11.85597 | 2.332444 |
| MR2 | 12.64357 | MR3 | 11.85597 | 0.787599 |
| MR4 | 8.86646 | MR3 | 11.85597 | 2.989514 |
| MR6 | 15.81331 | MR3 | 11.85597 | 3.957335 |
| MR7 | 18.0022 | MR3 | 11.85597 | 6.146224 |
| **MR8** | **12.51121** | **MR3** | **11.85597** | **0.655239** |
| MR9 | 8.997329 | MR3 | 11.85597 | 2.858645 |

**TABLE. XVII   DISTANCE BETWEEN MAPPING DISTANCE OF GROUP-1 MEDICAL RECORDS AND MR5**

| Records | Mapping Distance | Record | Mapping Distance | Distance |
|---------|-----------------|--------|-----------------|----------|
| MR1 | 9.52353 | MR3 | 11.86645 | 2.342919 |
| MR2 | 12.64357 | MR3 | 11.86645 | 0.777124 |
| MR4 | 8.86646 | MR3 | 11.86645 | 2.999989 |
| MR6 | 15.81331 | MR3 | 11.86645 | 3.946861 |
| MR7 | 18.0022 | MR3 | 11.86645 | 6.135749 |
| **MR8** | **12.51121** | **MR3** | **11.86645** | **0.644764** |
| MR9 | 8.997329 | MR3 | 11.86645 | 2.86912 |

From Table XVIII shows the attribute value imputed for attribute Z3 of record MR3 is 2. i.e the categorical attribute value $K_{32}$. This is because the attribute value, $K_{32}$ was mapped to numerical value **2.** From Table XIX shows the attribute value imputed for attribute Z4 of record MR5is 7. i.e NUMERIC value.

**TABLE.XVIII   IMPUTATION OF A3 ATTRIBUTE VALUE IN MR3**

| Medical Record | Z1 | Z2 | Z3 | Z4 | Decision Class |
|----------------|----|----|----|----|----------------|
| MR3 | 3 | 6 | 2 | 7 | CLASS-2 |

**TABLE.XIX IMPUTATION OF A4 ATTRIBUTE VALUE IN MR5**

| Medical Record | Z1 | Z2 | Z3 | Z4 | Decision Class |
|----------------|----|----|----|----|----------------|
| MR8 | 3 | 6 | 2 | 7 | CLASS-2 |

To fill the missing values for numerical attributes, we may impute the mean value of the corresponding record class identified using proposed approach. Alternately we may fill missing attribute values using several other strategies once we know nearest record.

## 5. CONCLUSIONS

The main contribution towards research in this paper is the approach adopted for fixing and estimating missing values in medical records. The idea is to perform imputation through class based clustering approach by clustering medical records to a number of classes equal to decision class labels. Also we try to find nearest neighbors for these medical records use the same to obtain final mapping distance. This is done for both missing value and non-missing value records. The dimensionality of medical records is also a major concern when trying to impute missing values and we achieve dimensionality reduction by using concept of mapping distance in this paper. The reduced dimension is a single dimension value. In essence, each record is transformed to a single dimension. To the best of our knowledge this approach is not used in the literature and this coins the importance of proposed approach. The case study demonstrates importance of proposed approach to impute both categorical and numeric values of medical records.